\documentclass[twocolumn]{article}
\usepackage[]{amsmath}
\usepackage[dvips]{epsfig}%
\begin{document}

\title{The PSI np data and their effect on  the  charged 
$\pi$NN coupling constant }
\author{ J. Franz, E. R\"ossle, H. Schmitt, and L. Schmitt \\
\small Fakult\"at f\"ur Physik der Universit\"at Freiburg 
D-79104 Freiburg, Federal Republic of Germany \\
 \small\em August 6, 1999 }
\date{\normalsize PACS Ref: 
      {13.75.Cs} {Nucleon-nucleon interaction}\hfill \mbox{ }
      }
\maketitle
\begin{abstract}\small
Differential cross sections of elastic neutron-proton scattering have been
measured for the energy range from 200 MeV to 580 MeV. The angular interval
for the detection of the recoiling proton ranges up to 48$^{\circ }$ in
the laboratory system. This corresponds to an interval of the scattering angle 
from 80$^{\circ }$ to 180$^{\circ }$ in the centre-of-mass system. For 
absolute normalization the simultaneously measured $\rm np \to d\pi^o$ 
reaction was used above 280 MeV. 
The charged $\rm\pi NN$ coupling constant has been determined to 
$f^2_{\rm\pi np} = 0.076 \pm 0.001$. 
\end{abstract}
\section{Introduction}
Neutron-proton elastic scattering at backward angles in the medium
energy regime has been the subject of several experiments\,[1--6].
\nocite{SHE74,BIZ75,BON78,Hue801,BUG82,NOR93}
The common feature of these investigations is a steep rise of the
differential cross section towards the back scattering angle of 180$^{\circ
} $. 
The slope of the sharp backward peak suggests a connection to the one pion
exchange (OPE) amplitude. 
Several suggestions have been made for a theoretical description 
of the experiments\,[7--16]. 
\nocite{PHI63,PHI67,HEN69,ROS70,KAN70,DIU73,DIU75,FER92,JAI92,GIB94} 
All of these proposals are able to describe the
backward spike, but most of them are purely phenomenological and some of
them fail to describe other observables. 

Precise backward scattering data offer the opportunity to evaluate the
pion--nucleon coupling constant $f^2_{\pi{\rm NN}}$\,\cite{Chew}. The medium energy
region is particularly suited for this purpose, because the pion pole is not
very far from the physical region, so that one could expect a reliable
extrapolation to it. However, the values determined so far by this method 
are not all in accordance with each other and with 
the value obtained from pion-nucleon scattering\,\cite{Hoehler}.

The results presented in this paper  have been obtained from four separate
experiments of different angular ranges, labeled I--IV\,[4, 19--21].
\nocite{Hue801,Franz99,Rup81di,Hab90di} 
Together they 
span the interval from about 80$^{\circ
}$ to 180$^{\circ }$ in the centre-of-mass system. 
The neutron energies range from 200~MeV to 580~MeV in steps of 20~MeV.

\section{Experiment} \label{expsetup}

The experimental set-ups and techniques of experiments I-IV have been
quite similar but with differences in detail of the accelerator performance,
the beam arrangement and the detection equipment. Here we outline the common
features only and refer to a forthcoming paper \,\cite{Franz99} for details. 

The experiments\, have been performed at the Paul Scherrer Institute (PSI), the
former Swiss Institute for Nuclear Research (SIN). The proton beam of the
ring cyclotron of 589 MeV energy consists of bunches with a width
of less than 1~ns at a rate of 50.63~MHz for experiments I and III and 
16.88~MHz for II and IV, corresponding to bunch spacings of 19.75~ns and 59.25~ns,
respectively. The beam current during the data taking was  60--100 ${
\mu}$A. The neutrons were produced on a thick target of 
beryllium (I--III) or carbon (IV). Neutrons escaped through a collimator 
hole in the beam dump at an angle of 60 mrad with respect to the 
incident protons. They were shaped by two additional 
collimators to a beam of about 2 x 2 cm$^2$ at a distance of 61 m from the 
production target. Charged particle
contaminations were eliminated by cleaning magnets behind the collimators.
A lead filter was inserted in order to reduce the 
photon component of the beam which mainly originates from the decay of 
neutral pions in the production target.

The continuous neutron energy spectrum consists of the 40~MeV wide 
quasielastic peak 
at about 540~MeV and a broad distribution at lower
energies,  which can be ascribed to inelastic processes \,\cite{Arnold98a}. 
The detailed shape of the spectrum depends on the target
material and its thickness.

After a flight path of 61 meters, kept at rough vacuum of about 100 Pa, 
the neutron beam hits a liquid hydrogen target. 
For the analysis of the scattering products a magnet spectrometer 
was installed on a turn table. The spectrometer was equipped with 
drift chambers and two scintillation 
counters, a thin one (1~mm thick) in front of the magnet, and a thicker 
one (1~cm thick) behind it. They allow to measure both, 
the time-of-flight of the 
particle detected by the spectrometer, and that of the incoming neutron with 
respect to the rf-signal of the accelerator. 
The angular acceptance is almost 20$^{\circ}$. The average momentum 
resolution is about 3 \% {\sc FWHM.}

\section{Data taking and analysis}
The data have been taken at different run periods, each extending over several 
weeks. The contribution of the target surroundings and spectrometer materials 
was measured with an empty target. It was subtracted after normalization to 
the neutron intensity of the full target measurement. Corrections have been applied 
for the event rate dependent dead-time losses.

The long periods of data taking required special attention to the 
long term variations of the whole system, particularly of the neutron 
time-of-flight measurement. The stability of the 
electronics has been checked regularly and close control
of the drift chamber gas flow and high voltage supply has been maintained.

The neutron intensity was monitored in three different ways. 
The integrated 
primary proton beam intensity was provided from the 
accelerator control centre. 
This signal, though convenient, was not sufficiently 
reliable because of variations 
of the focusing of the primary beam on the production target. 
The second monitor consisted of a three stage scintillation 
counter telescope, which  recorded charged
particles emerging from a 
thin polyethylene target placed in
the beam at about 44~m from the neutron target. 
A third monitor was installed behind the magnet
spectrometer. It recorded elastic np scattering events from a 
polyethylene block in coincidence. 

These monitors were intended to measure the relative
neutron intensity. Above 280~MeV the absolute normalization was performed by 
the simultaneously recorded $\rm np\to d\pi^o$ reaction as reference
cross section. This is discussed in sect.~\ref{dpinorm}.

\subsection{Time-of-flight calibration}

As mentioned above, the energy 
spectrum of the incident neutrons is continuous. Therefore, 
the neutron time--of--flight measurement is the basis of the incident 
energy determination, and the control of its stability is crucial. 
It requires careful calibration of the zero point and the conversion
gain of the time--to--digital converters. For the control of 
the time zero point we used the high
energy photons in the beam. The
lead filter in the neutron beam was removed for these calibration runs, and
the hydrogen target was replaced by a lead slab for a higher conversion rate.
With the spectrometer field inverted and reduced appropriately, the converted  
electrons  were detected. From the 
width of the peak the time resolution has been evaluated to be 0.8 ns {
FWHM} including the bunch width of the primary beam and the contribution 
from electronics. The corresponding neutron energy resolutions vary from 
1.2~MeV at 200~MeV to 7.6~MeV at 580~MeV. 

These calibrations have been performed in regular intervals.
Deviations from the overall mean have been corrected for each run.

\subsection{Data reduction and event selection}

The first step in the off-line analysis was a reduction of the data by
setting cuts in order to select good events. These cuts included a unique
track in the drift chambers, starting in the target volume and emitted
within the full acceptance of the spectrometer in vertical and horizontal
direction. 
Next, the mass of the particle in the spectrometer was determined 
from the measured time-of-flight through the spectrometer and the measured 
momentum. A mass selection has been applied by rejecting masses
$ m < 0.5\ m{\rm _p}$ and $ m >1.5\ m{\rm _p}$, where $m_{{\rm p}}$ is the
proton mass. The upper limit was increased to $ m >2.5\ m{\rm _p}$ 
for experiments I and II, in order to keep also deuteron events,   
which were used for the normalization of the np data 
(see sect. \ref{dpinorm}). Finally, low momentum particles have
been excluded by cuts at $p = 500$~MeV/c for protons and $p = 620$~MeV/c for
deuterons. 

\subsection{Neutron energy determination}

\label{ambig}
The emission angle of the recoil proton is determined by the
set of drift chambers in front of the magnet, and its momentum by the
deflection in the magnetic field. The proton energy is obtained after
correction of the energy loss in traversing the material of target and
spectrometer.  The incident neutron 
energy is determined from the measured 
neutron time-of-flight $t_{\ell}$. A complication arises from the fact, 
that the bunch interval given by the rf-signal (19.75 ns or 59.25 ns) 
is much shorter than 
the time-of-flight from the production target to the hydrogen target 
(more than 200 ns). This introduces ambiguities on the proper multiple of the 
bunch spacing time which has to be added to the measured value.  However, 
by comparison with the time-of-flight $t_{\rm calc,\ell}$, calculated from 
the measured recoil proton momentum vector, the correct number of bunch 
spacings can be determined.

\section{Experimental results}

\subsection{np differential cross sections}\label{polkorr}

The data were binned in energy intervals of 20 MeV, and angular bins of 0.5 
degrees in the centre-of-mass (CM) system. Each bin has been corrected for 
absorption of the recoil protons in the target and the spectrometer material. 
This energy dependent correction was below 1\,\% in all cases. Another 
correction was necessary due to a small polarization component of the 
neutron beam in vertical direction of $.05 < |P_{\rm n}| < .08$. 
This comes about since the neutrons 
are produced at an angle of 60~mrad (cf. sect.~\ref{expsetup}). 
The energy and angle dependent effect of the analyzing power of the np 
elastic scattering has been taken into account. 
The correction to the differential cross section was below 2.5\,\%. The 
errors of these corrections contribute to the 
errors of the differential cross sections with less than 0.3\,\%.
\begin{figure*}[tb]
\centerline{\epsfig{file=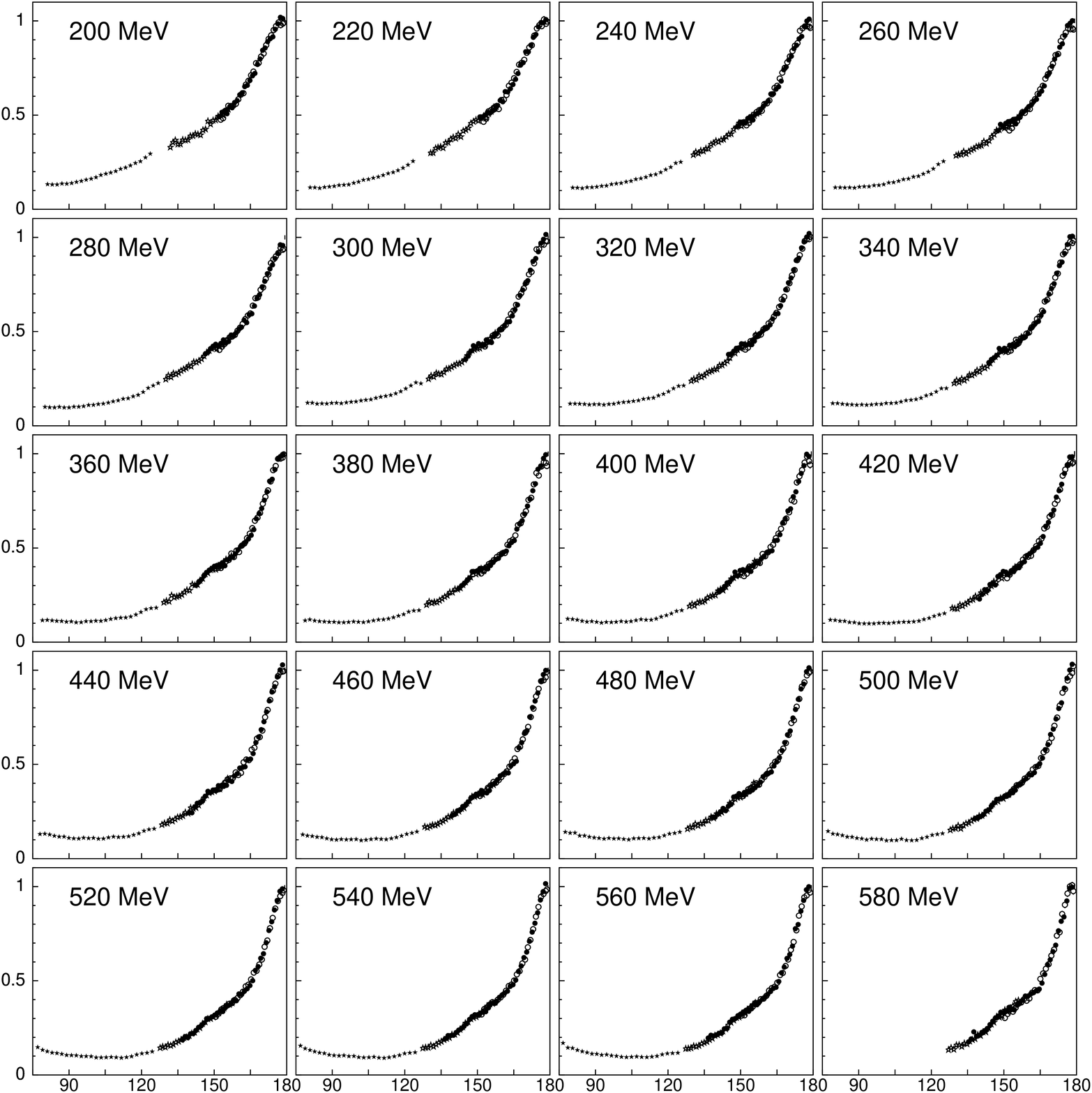,width=18cm,height=16cm}}
\caption{\label{Franzfig1}\small Relative differential np cross sections as a function of the 
CM-scattering angle $\theta_{\rm CM}$. 
$\circ$:~data set~I, $150^{\circ} \le \theta_{\rm CM} \le 180^{\circ}$; $\bullet$:~data 
set~II, $140^{\circ} \le \theta_{\rm CM} \le 180^{\circ}$; $\times$:~data set~III, 
$129^{\circ} \le \theta_{\rm CM} \le 156^{\circ}$; 
$\star$:~data set~IV, $80^{\circ} \le \theta_{\rm CM} \le 126^{\circ}$. 
The data are normalized to 1 at $\theta_{\rm CM} = 179.2^\circ$. 
Errors are not visible within the symbol size.
}
\end{figure*}

The four sets of data have been combined in an appropriate way. For the
experiments I and II a common angular region exists which has been used to
evaluate the multiplying factor for experiment I. The uncertainty introduced
by this procedure for the different energies has been determined as 0.6\,\% from the
spread of the individual factors from the average.
In a similar way the data set from 
experiment III has been linked to the combined data of set I and set II. 
Here the uncertainty was 0.7\,\%. 
This procedure could not be followed for the normalization of experiment IV
because no overlap of data points exists. We therefore fitted for each
neutron energy the data points of experiment III for CM-angles smaller than
145$^\circ$ together with the same number of data points of set IV by 
Legendre polynomials of 4th 
order with a free scaling factor. Each data point has been given the same
statistical weight of 2 \% in the fit procedure in order to assign the same
weight for the two data sets. The average value of $\chi ^2$ for the
relative normalization was 1.41, and the mean error of the scaling factor is 2\,\%.

The statistical errors of the differential cross sections 
range from 1\,\% to 2\,\%. The spread of the data with respect to 
smooth fits with polynomials of an appropriate order is more like 2--2.5\,\%. 
This reflects systematic uncertainties due to 
\begin{itemize}
\item[-] short term changes in the phase of the rf-signal
\item[-] imperfections of the drift chambers and other detectors
\item[-] remaining effects of gain shifts of detectors and electronics.
\end{itemize}
Taken together they are in the order of 1--2\,\%, depending on the set-up. 
Systematic errors of 1.4\,\%, 0.6\,\% and 1.7\,\% have been added in 
quadrature to the data sets II, III and IV, respectively. A systematic 
error of 0.5\,\% had been added earlier to data set I \,\cite{Hue78do}.

The relative differential cross sections for the four separate experiments are
displayed together in Fig.~\ref{Franzfig1}. The angular distributions are normalized 
to 1.0 at the largest angle of data set II. The numerical values are tabulated 
in ref. \,\cite{FRE99}.

\subsection{Discussion}
A common feature of the angular dependence at all energies is the sharp rise
towards the backward direction. It is followed by a less steep decrease
passing through a wide minimum, which is shifted to smaller angles with
decreasing energy. In the transition region of the two different slopes
there is an indication of a bump. A
similar shape is indicated in the LAMPF data \,\cite{BON78} but not as clear.

\begin{figure}[h]
\vspace*{-.7cm}
\centerline{\epsfig{file=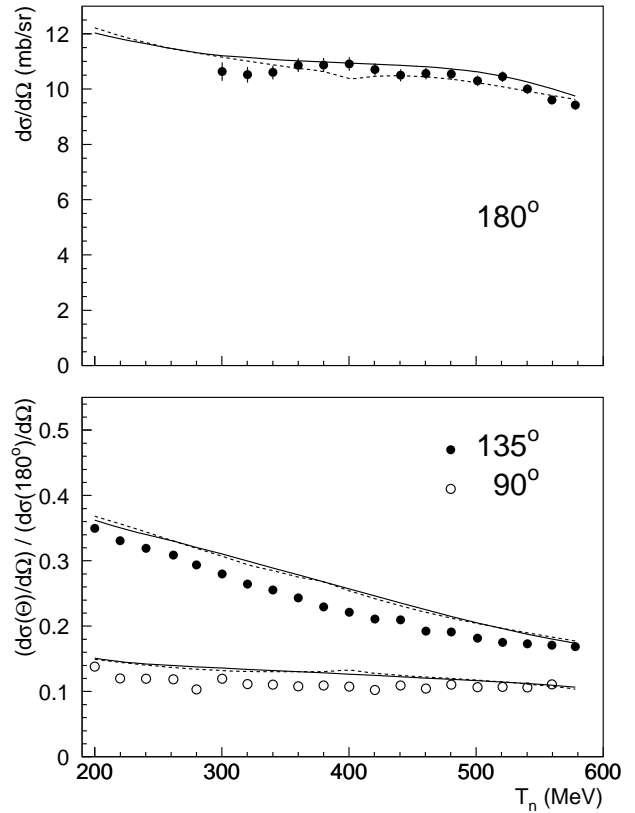,width=8.5cm}}
\caption{\label{Franzfig2} \small
Comparison of our data with phase shift predictions. Full lines: 
Arndt et al. \,\cite{ARN95}; 
dotted lines: Bystricky et al. \,\cite{BYS}. In the upper diagram the cross sections 
extrapolated to $180^\circ$ are shown. In the lower diagram the cross section ratios 
$\frac{\rm d\sigma}{\rm d\Omega}(\theta)/
\frac{\rm d\sigma}{\rm d\Omega}(180^{\circ })$ vs. $T_{\rm n}$ are shown for 
$\theta = 135^\circ$ and $\theta = 90^\circ$.
\hfill {\ }}
\end{figure}

The energy dependence of the cross section
at 180$^{\circ }$ is shown in Fig.~\ref{Franzfig2}. 
It is roughly constant in our energy range, and is well described by the 
phase shift predictions.
With decreasing angle a deviation of the data from the phase shift
predictions develops within the full energy range. 
This can be clearly seen by a
comparison of the cross section ratios $\sigma (\theta )/\sigma (180^{\circ
})$ as shown in Fig.~\ref{Franzfig2} 
for $\theta $ = 135$^{\circ }$ and 90$^{\circ }$. The
deviations occur for both phase shift solutions of Arndt et al. \,\cite{ARN95}
and Bystricky et al. \,\cite{BYS}. It reflects the fact that for the extreme
backward scattering angles several precise and consistent measurements exist 
\,\cite{BON78} which pin down the phase shift solutions, whereas the angular
region around $\theta  = 90^{\circ }$ has been covered only scarcely at
singular energies so far. 
The cross section ratios, as displayed in Fig.~\ref{Franzfig2}, are decreasing with
increasing energy, except for the ratio $\sigma (90^{\circ })/\sigma
(180^{\circ })$ which is almost energy independent above 250 MeV.

\section{Absolute normalization}
\label{dpinorm}
\label{nocor}

For the absolute normalization of the cross section the 
incident neutron 
intensity has to be known as a function of energy. 
In our case the neutron
intensity was obtained for experiment II from a comparison with 
the simultaneously measured reaction
\begin{equation}
{\rm np\rightarrow d\pi ^0\;,}\label{npdpi0}
\end{equation}  
where the deuterons have been recorded by the spectrometer like the 
recoiling protons of the elastic scattering. 

Isospin independence is used to relate the  
cross section of the process (\ref{npdpi0}) to
the cross section of  reaction 
\begin{equation}
{\rm pp\rightarrow d\pi ^+}\;,\label{ppdpip}
\end{equation}
for which precise data exist. 
However, isospin symmetry, which relates the two reactions by 
$\sigma_{\rm d\pi^o} = \frac{1}{2} \sigma_{\rm d\pi^+}$ 
is not exact since the masses in the 
initial and in the final states are different for (\ref{npdpi0}) and 
(\ref{ppdpip}). Also,  
the Coulomb interaction for  process (\ref{ppdpip})
must be taken into account. 
This is discussed in detail in section \ref{coulomb}.

\subsection{\boldmath Measurement of $\rm \frac{ d\sigma}{d\Omega}
(np\to d\pi^o)$}

 The angular acceptance of the magnet spectrometer is 
sufficiently large to cover the full angular range of the deuterons, 
$\theta_{\rm d}^{\rm max}\le 12^o$, so that 
a complete angular distribution of reaction (\ref{npdpi0}) is obtained. 
The energy of the deuterons emitted in (\ref{npdpi0}) varies strongly with the emission angle. For an incident neutron
energy of 400 MeV,  as an example, the deuteron energy varies from 
130 MeV to 260 MeV. In this energy region the deuteron cross section on nuclei is
strongly energy dependent and distorts the angular distribution. Besides the
energy loss the absorption by the spectrometer material has to be corrected
for. While the energy loss correction is straightforward, the correction for
absorption loss is more involved because of the missing knowledge of
deuteron cross sections on nuclei.

For the deuteron total cross section on a nucleus with
mass number $A$ we use \,\cite{WIL71}
\begin{equation}\label{equat3}
\sigma_{{\rm d}A} = 0.97 (\sigma_{{\rm n}A} + \sigma _{{\rm p}A}) \;.
\end{equation}
The material around the target and the spectrometer consists
mainly of hydrogen and the self-conjugate nuclei C, N, O and Ar, 
for which we assume $\sigma _{{\rm n}A}=\sigma _{
{\rm p}A}$. The deuteron cross section eq. (\ref{equat3}) reduces then to \fussy
\begin{equation}
\sigma_{{\rm d}A} = 1.94 \sigma_{{\rm n}A}\;.
\end{equation}
Neutron total cross sections on nuclei have been measured in a previous
work \,\cite{FRA88}. For N and Ar, which have not been measured
directly, the cross section has been evaluated from a parameterization of the
energy and mass number dependence given in \,\cite{FRA88}.
For the deuteron cross section on hydrogen $\sigma _{{\rm dp}}$ at the energy $T$
we used $\sigma _{{\rm pd}}$ data or the   $\sigma _{
{\rm np}}$ data of \,\cite{Gru85}  at $T/2$.

The absorption correction is then obtained by summing up the contributions
of the different elements according to their thickness in the spectrometer.
The resulting correction varies from 5.2\,\% at $T{\rm _d} = 120$~MeV to 
2.5\,\% at 400 MeV. As in the case of elastic np scattering (cf. sect. \ref{polkorr}) 
the effect of a small polarization of the incoming neutrons has been corrected for. 
This angular and energy dependent correction for the deuteron 
production is below 0.6\,\% for $T_{\rm n} < 490$~MeV. 
At our highest energy it reaches 2.6\,\%. The errors of both corrections 
taken together contribute to the errors of the differential production cross 
sections with less than 0.5\,\%.

\subsection{ Mass- and Coulomb corrections }
\label{coulomb}
\label{normdpi}

Niskanen and Vestama \,\cite{nv} studied the symmetry violating effects for 
the reactions (\ref{npdpi0}) and (\ref{ppdpip}) in 
the framework of a coupled-channel me\-thod with different pion production 
mechanisms. Because the final momentum is most relevant in threshold reactions 
with a large negative Q-value 
the comparisons are made at the same final pion momenta $p_{\pi}$ in the CM system, 
commonly expressed by the dimensionless variable $\eta = p_{\pi}/m_{\pi^+}$. 
For both reactions, the pion production cross section can be factorized as 
${\rm d}\sigma /{\rm  d}\Omega = P \cdot R $ with the phase space 
factor $P$ and $R$ the sum of the squared matrix elements of the pion 
production operator. With this ansatz the relative change in the 
cross sections was determined in good approximation as
\begin{equation}
\frac{\delta \sigma }{ \sigma_{\rm av}}  \approx  \frac{\delta P}{P_{\rm av}} + 
\frac{\delta R }{ R_{\rm av}}
\end{equation}
with
\[ \begin{array}[h]{rlcrl}
\delta \sigma   = & 2\,\sigma_{{\rm d}\pi^o}  -  \sigma_{{\rm d}\pi^+}  & \; & 
\sigma_{\rm av}  = & 
{\textstyle\frac{1}{2}}\cdot(2 \sigma_{{\rm d}\pi^o}  +  
\sigma_{{\rm d}\pi^+}) \nonumber \\[2mm]
\delta R  = & R_{{\rm d}\pi^o}  -  R_{{\rm d}\pi^+}  & \; & 
R_{\rm av}  = & {\textstyle\frac{1}{2}}\cdot( R_{{\rm d}\pi^o}  + 
 R_{{\rm d}\pi^+}) \nonumber \\[2mm]
\delta P   = & P_{{\rm d}\pi^o}  -  P_{{\rm d}\pi^+}  & \; & 
P_{\rm av}  = & {\textstyle\frac{1}{2}}\cdot( P_{{\rm d}\pi^o}  + 
 P_{{\rm d}\pi^+}) \;. \nonumber \\ 
\end{array} %
\] %
Niskanen and Vestama calculated  the effect due to the 
mass differences 
and incorporated the Cou\-lomb correction 
in the final state only. At higher proton energies, 
the Coulomb correction for the pp initial state is 
comparable with the one in the d$\pi^+$ final state. Therefore we use the 
results of Ref.~\,\cite{nv} for the change in the matrix element without 
Coulomb correction and apply Coulomb penetration factors for a point proton 
with an extended proton and for a point pion with an extended deuteron. 
The weighted sum of the Coulomb penetration factors 
for the different orbital angular 
momentum states with weighting factors from \,\cite{ARN2} is used as 
Coulomb correction $C^2_{ec}$
for the squared matrix elements $R_{0\rm{d}\pi^+}$ without Coulomb correction, 
\begin{equation}
R_{\rm{d}\pi^+} = C^2_{ec} \cdot R_{0\rm{d}\pi^+}\,. \label{Cecq}
\end{equation}
Table~\ref{Franztab1} shows some relevant information 
used to 
transform the cross section $\sigma_{\rm d\pi^+}$ of (\ref{ppdpip}) to 
$\sigma_{\rm d\pi^o}$ of the 
process (\ref{npdpi0}) by the relation
\begin{equation}
\sigma_{\rm d\pi^0} = \frac{1}{2} \, \sigma_{\rm d\pi^+} \,\cdot\,\frac{ 2 + 
\delta \sigma / \sigma_{\rm av}} { 2 - \delta \sigma / \sigma_{\rm av}}.
 \end{equation}
\begin{table*}
\begin{center}
\begin{tabular}{|c|c|c|c|c|c|c|c|ccc|ccc|}
\hline
1 & 2 & 3 & 4 & 5 & 6 & 7 & 8 &  \multicolumn{3}{c|}{ 9 } & \multicolumn{3}{c|}{10} \\
\hline \rule[3.5ex]{0pt}{0pt} 
$T_{\rm n}$ & $\eta$ &  $T_{\rm{p}}$ & $C_{ec}^2$ & $\frac{\delta R}{R_{\rm av}}$ & 
$\frac{\delta P}{P_{\rm av}}$ & $\frac{\delta \sigma}{\sigma_{\rm av}}$ & 
$\sigma_{\rm d\pi^+}$ & 
$\sigma_{\rm d\pi^o}$  &  & $\Delta\sigma$   & 
$\frac{{\rm d}\sigma}{{\rm d}\Omega}(\theta_{\rm max})_{\rm np}$ &  
& $\Delta\frac{{\rm d}\sigma}{{\rm d}\Omega}$ \\ 
{(MeV)}\rule[2.5ex]{0pt}{0mm}   &   & (MeV) &  &  &  &  & (mb) &  (mb) &  & (\%) & 
(mb/sr) &  & (\%) 
\\   \hline
300.2 & 0.3960  & 312.0 & 0.9364 &0.1194 & -0.0095 & 0.1099 & 0.1225 & 0.0684 & $\pm$ & 2.6 & 10.63 & $\pm$ & 3.1 \\
320.1 & 0.5369  & 331.4 & 0.9471 &0.0950 & -0.0084 & 0.0866 & 0.2288 & 0.1248 & $\pm$ & 2.2 & 10.52 & $\pm$ & 2.7 \\
340.0 & 0.6527  & 350.9 & 0.9525 &0.0778 & -0.0074 & 0.0704 & 0.3639 & 0.1952 & $\pm$ & 2.1 & 10.60 & $\pm$ & 2.4 \\
360.1 & 0.7558  & 370.7 & 0.9559 &0.0643 & -0.0066 & 0.0577 & 0.5296 & 0.2805 & $\pm$ & 2.1 & 10.86 & $\pm$ & 2.3 \\
379.9 & 0.8486  & 390.1 & 0.9584 &0.0521 & -0.0059 & 0.0462 & 0.7213 & 0.3777 & $\pm$ & 2.0 & 10.87 & $\pm$ & 2.3 \\
400.2 & 0.9375  & 410.1 & 0.9606 &0.0412 & -0.0053 & 0.0359 & 0.9474 & 0.4910 & $\pm$ & 1.9 & 10.91 & $\pm$ & 2.2 \\
420.3 & 1.0206  & 429.9 & 0.9622 &0.0325 & -0.0048 & 0.0277 & 1.200  & 0.6171 & $\pm$ & 1.9 & 10.71 & $\pm$ & 2.1 \\
440.5 & 1.1004  & 449.9 & 0.9638 &0.0254 & -0.0043 & 0.0211 & 1.481  & 0.7563 & $\pm$ & 1.8 & 10.50 & $\pm$ & 2.1 \\
460.4 & 1.1760  & 469.6 & 0.9651 &0.0200 & -0.0039 & 0.0161 & 1.777  & 0.9030 & $\pm$ & 1.8 & 10.56 & $\pm$ & 2.0 \\
480.5 & 1.2498  & 489.5 & 0.9665 &0.0146 & -0.0036 & 0.0110 & 2.085  & 1.054  & $\pm$ & 1.8 & 10.54 & $\pm$ & 2.0 \\
501.0 & 1.3229  & 509.8 & 0.9678 &0.0132 & -0.0032 & 0.0100 & 2.392  & 1.208  & $\pm$ & 1.7 & 10.30 & $\pm$ & 1.8 \\ 
520.8 & 1.3916  & 529.4 & 0.9690 &0.0225 & -0.0030 & 0.0195 & 2.664  & 1.358  & $\pm$ & 1.7 & 10.45 & $\pm$ & 1.8 \\
540.3 & 1.4577  & 548.7 & 0.9700 &0.0400 & -0.0027 & 0.0373 & 2.889  & 1.499  & $\pm$ & 1.7 & 10.00 & $\pm$ & 1.8 \\
559.6 & 1.5217  & 567.9 & 0.9711 &0.0560 & -0.0025 & 0.0535 & 3.054  & 1.611  & $\pm$ & 1.7 & 9.61  & $\pm$ & 1.8 \\
578.1 & 1.5820  & 586.2 & 0.9719 &0.0700 & -0.0023 & 0.0677 & 3.143  & 1.682  & $\pm$ & 1.7 & 9.42  & $\pm$ & 1.9 \\
  \hline
  \end{tabular}
\caption{\label{Franztab1} \small Absolute normalization of the elastic np  
cross sections. All cross section related quantities are given in the CM system. 
Col.~1: incident neutron energy; 
col.~2: $\eta = p_{\pi}/m_{\pi^+}$ for reaction \ref{npdpi0}; 
col.~3:  incident proton energy for reaction \ref{ppdpip} at same $\eta$; 
col.~4--7: Contributions to Mass and Coulomb correction (details see text);
col.~8: integrated cross section of reaction (\ref{ppdpip}); 
col.~9: integrated cross section of reaction (\ref{npdpi0}); 
col.~10: absolute normalization factor for 
np elastic cross sections}
\end{center}
\end{table*}
The uncertainty for $\delta R /R_{\rm av}$ is according to Ref.~\,\cite{nv} 
about 0.01. Together with the Coulomb corrections which we applied to the pp initial 
state, this increases to 0.015. This error contribution is 
included in the errors given in col.~9 of Table~\ref{Franztab1}.

\subsection{\boldmath Parameterization of $\sigma(\rm pp \to d\pi^+$)}

Cross sections for reaction (\ref{ppdpip})
as well as for the inverse reaction
$\rm \pi^+ d \rightarrow pp $ 
have been measured with sufficient accuracy in the relevant energy range from 
threshold up to 640 MeV. 
The  data base [31--35] \nocite{ARN2,neuDat,neuDat2,neuDat3,neuDat4} consists of 61 integrated 
cross sections for the reaction (\ref{ppdpip}) between 
288~and~641~MeV incident proton energy, and of 74 cross sections for 
the inverse process 
between 1.8~and~174~MeV incident pion energy. 
The latter ones have been converted to  (\ref{ppdpip})
via detailed balance. Statistical and systematic errors of each data point 
have been  added in quadrature (Gaussian). 
The 
data set has been fitted by a parameterization similar to Bystricky et al. 
\,\cite{BYS2} as a function of $\eta$. The fit error of $\sigma_{\rm d\pi^+}$ 
ranges from 2.1\,\% at $\eta = 0.396$ to 0.7\,\% at $\eta = 1.582$.

\subsection{Absolute normalization above 300\,MeV}

The absolute normalization factor for the elastic np scattering data 
is given with respect to the highest measured angle $\theta_{\rm max} = 179.2^\circ$ of data set II
\begin{equation}
 \frac{{\rm d}\sigma}{{\rm d}\Omega}(\theta_{\rm max})_{\rm np} = \frac{\tilde{N}_{\rm 
 np}(\theta_{\rm max})}{\rm 
 \int_{}^{}\tilde{N}_{{\rm d}\pi^0} {\rm d}\Omega_{\rm 
 d\pi^o}}\cdot\sigma_{\rm d\pi^o}\;, \label{nprate}
\end{equation}
where $\tilde{N}_{\rm np}$ and $\tilde{N}_{d\pi^0}$ are the solid angle corrected event 
numbers for the two processes in corresponding energy bins.
The results are given in  col.~10 of Table~\ref{Franztab1}. 
Close to pion threshold and below, 
this method can not be applied. 

\section{\boldmath The $\pi\rm N N$ coupling constant $f_{\rm\pi NN}^2$ }
\label{fq}
\subsection{The Chew method}
The value of the ${\pi}NN$ coupling constant $f^2_{\rm\pi NN}$ has focussed 
new interest recently in
connection with a discussion on a conceivable breaking of charge 
independence\,[37--41].\nocite{fqArndt,fqSwart,Stoks,fqEricson,fqBugg} The 
charged coupling constant, $f^2_{\rm c}$, can be obtained by an extrapolation of the
measured backward elastic differential cross section to the pion pole.  The standard procedure in the past has been the 
Chew method \,\cite{NOR93,Chew,Cziffra}, which extrapolates the 
Chew funcion $y(x)$ defined by
\begin{equation}
\label{feq1}
y(x) = (x - x_{\pi -pole})^2 \cdot \frac{\rm{d}\sigma}{\rm{d}\Omega} = 
\sum_{j=0}^{j_{\rm max}} a_j P_j(x) 
\end{equation}
to the pion pole in the unphysical region.
Here $x = \cos{\theta}$ with $\theta$ the CM-angle of the scattered neutron 
and 
\begin{eqnarray}
\label{feq2}
x_{\pi -pole} & = & \cos{\theta_{\pi-pole}} \nonumber \\ 
 & = & 
- (m_{\pi}^2 - m_{\rm n}^2 - m_{\rm p}^2 + 2E_{\rm n}E_{\rm p})/2p_{\rm n}^2
\end{eqnarray}
with $m_{\pi}$, $m_{\rm n}$ and $m_{\rm p}$ the masses of the charged pion, the
neutron and the proton, $E_{\rm n}$ and $E_{\rm p}$ the CM-energies of 
the neutron and the proton and $p_{\rm n}$ the CM-momentum of the neutron.
The $P_j$ are polynomials generated with special recurrence relations, 
so that they are orthogonal in the range where data points
exist \,\cite{Hamm}. At the pion pole the Chew function gives
\begin{eqnarray}
     y(x_{\pi-pole}) & = & \sum_{j=0}^{j_{\rm max}} a_j P_j(x_{\pi-pole}) 
     \nonumber \\
 & = & (\hbar {\rm c})^2 f_{\rm c}^4 (m_{\rm n} + m_{\rm p})^4 / (4 s 
 p_{\rm n}^4)
\end{eqnarray}
where $s$ is the total energy squared.
\subsection{Test with pseudo data}\label{pseudat}
We have tested this extraction method with pseudo data between 240 MeV and
540 MeV neutron laboratory kinetic energy in 100 MeV steps. The data were 
generated from the regularized OPE model of Gibbs and Loiseau \,\cite{GIB94} 
in the angular range from $80^\circ$ to $180^\circ$  in $1^\circ$ 
steps. The input coupling constant
was fixed at $f_{\rm c}^2 = 0.076$. Uncertainties of the pseudo data were 
generated with a
\begin{figure}[h]
\centerline{\epsfig{file=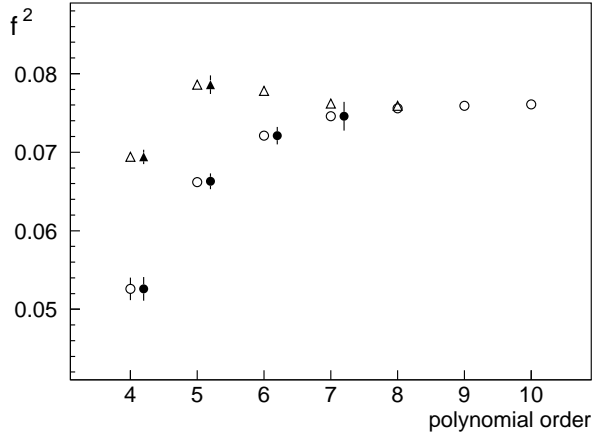,width=8.5cm}}
\caption{\label{Franzfig3} \small $f^2$ fits at $T_{\rm n} = 440 {\rm MeV}$ vs. 
maximum polynomial order.
Circles represent fits without, triangles  
with conformal mapping. Filled symbols are used for pseudo data with 
realistic errors (2\,\%), whereas open symbols are used for high precision 
pseudo data (0.1\,\%).}
\end{figure} 
Gaussian random error distribution. If these uncertainties are 
$\leq  0.1\,\%$, the model coupling constant can be  
reproduced with an 
error  of $\approx 0.25$\,\%. 
For pseudo data with uncertainties of 2\,\%, corresponding to the present experiment, 
the error for $f_{\rm c}^2$  is 
2.4\,\%. Moreover, the extracted
coupling constant is systematically smaller than its input value. 
This is because the extracted $f_{\rm c}^2$
rises with the maximum polynomial order of the fit as shown in 
Fig.~\ref{Franzfig3} (open circles). 
Applying an $F$ test \,\cite{Bev} to the fit 
gives a lower maximum order of the polynomials of eq. \ref{feq1}
for data with higher uncertainties.
At $T_{\rm n} = 440 $ MeV, for instance, and for uncertainties 
$\leq  0.1\,\%$ the $F$ test results in the maximum order 
$j_{\rm max} = 10$
and $f_{\rm c}^2 = 0.0761 \pm 0.2 \,\% $. For uncertainties of $2\%$ the 
$F$ test results in $j_{\rm max} = 7$ and $f_{\rm c}^2 = 0.0746 \pm 2.4 \,\%$.
\subsection{Chew method with conformal mapping}
The convergence properties of the polynomial expansion is improved by  conformal
mapping of the variable $x$ before extrapolating to the pole \,\cite{Loch}. 
We used this method similar to 
Dumbrajs et al. \,\cite{Dum84}. The position of the  neutral pion pole and the 
onset of the two pion ($\pi^o \pi^+$) exchange cut are mapped to (+1,0) 
and (-1,0) in the complex plane, respectively. 
The other exchange cuts lie on the unit circle. In addition, 
the data points are symmetrized around zero.

This mapping method was tested as before with pseudo data. 
Again, the model coupling constant is
reproduced for uncertainties  $\leq  0.1$\,\%. 
The convergence is faster than without 
conformal mapping, and the extrapolation error 
is of the same order or slightly larger. For the example at $T_{\rm n} 
= 440$~MeV given above, 
the $F$~test results a value of 
$f_{ \rm c,\,map}^2 = 0.0759 \pm 0.3\, \%$
with a reduced polynomial order of $j_{\rm max} = 8$ 
for uncertainties of $\le 0.1\,\%$. 
With uncertainties of 2\,\%, the 
$F$ test results in the lower value $j_{\rm max} = 5$ and 
$f_{\rm c,\,map}^2 = 0.0786 \pm 1.6\,\%$. Thus, the conformal mapping method results 
in a systematic upward 
shift for the extracted coupling constant, as shown by the triangles 
in Fig.~\ref{Franzfig3}. 
Again, this comes about because the 
extracted $f_{\rm c,\, map}^2$ depends on the maximum polynomial order, which is 
lower for less precise pseudo data.

 Similar results were obtained with pseudo data from phase shift predictions 
of Arndt et al. \,\cite{ARN95}. Again,  
the input value $f_{\rm c}^2 = 0.076$ could be  
reproduced only with high precision
($\leq  0.1$\,\%) pseudo data by both methods. For
uncertainties of 2\,\% the result was 
systematically too low without, and  too high with conformal mapping.  
Thus, both methods have systematic offsets of opposite sign. In principle, they 
can be taken into account by a correction factor $k_{\rm corr} = 
f^2_{\rm input}/f^2_{\rm extracted}$. 
 
\begin{table*}[hbt]
\begin{center}
\begin{tabular}{|c|c|c|c|c|c|c|c|c|c|c|c|}
\hline 
\multicolumn{4}{|c|}{\rule[3.5mm]{0pt}{0pt}} & 
\multicolumn{8}{c|}{results of Chew extrapolation}  
\\
\multicolumn{4}{|c|}{\raisebox{1.5ex}[-1.5ex]{ input  data }} & 
\multicolumn{4}{c|}{without conformal mapping} & 
\multicolumn{4}{c|}{with conformal mapping} \\
\hline \rule[3.5mm]{0pt}{0pt}
$T_{\rm{n}}$ & $\Delta\theta $ & $N$ & $s$ & 
$j_{\rm max}$ &  $f_{\rm c}^2 $ & $ \Delta f^2 $ & $k_{\rm corr}$  &
$j_{\rm max}$ &  $f_{\rm c,\,map}^2 $ & $ \Delta f^2 $ & $k_{\rm corr}$ 
\\ 
(MeV) & $({}^\circ)$  & &$(\%)$ & & & $(\%)$  & &  & & $(\%)$ &  \rule[-.5mm]{0pt}{0pt}
\\  \hline
 300.2 & 94.2 & 108 & 2.3 & 7 &  0.0766 & 4.7 & 0.980 & 5 &  0.0730 & 3.1 & 0.977 \\
 320.1 & 92.5 & 106 & 2.3 & 7 &  0.0778 & 4.2 & 0.983 & 5 &  0.0762 & 2.8 & 0.975 \\
 340.0 & 88.8 & 106 & 2.4 & 6 &  0.0713 & 3.4 & 1.040 & 5 &  0.0765 & 3.1 & 0.973 \\
 360.1 & 87.0 & 106 & 2.4 & 7 &  0.0794 & 4.4 & 0.985 & 5 &  0.0807 & 3.2 & 0.974 \\
 379.9 & 85.2 & 104 & 2.5 & 7 &  0.0791 & 4.5 & 0.982 & 5 &  0.0778 & 3.2 & 0.971 \\
 400.2 & 83.5 & 104 & 2.8 & 6 &  0.0713 & 3.9 & 1.049 & 5 &  0.0786 & 3.6 & 0.969 \\
 420.3 & 83.7 & 106 & 2.8 & 6 &  0.0700 & 3.8 & 1.056 & 5 &  0.0776 & 3.4 & 0.970 \\
 440.5 & 79.9 & 103 & 2.8 & 6 &  0.0742 & 3.5 & 1.049 & 5 &  0.0797 & 3.6 & 0.967 \\
 460.4 & 80.1 & 104 & 2.5 & 6 &  0.0725 & 3.1 & 1.060 & 5 &  0.0780 & 3.1 & 0.967 \\
 480.5 & 78.3 & 103 & 2.6 & 6 &  0.0734 & 3.2 & 1.060 & 5 &  0.0792 & 3.2 & 0.965 \\
 501.0 & 74.4 & 101 & 2.5 & 6 &  0.0757 & 3.0 & 1.053 & 5 &  0.0783 & 3.3 & 0.965 \\
 520.8 & 74.6 & 104 & 2.1 & 6 &  0.0769 & 2.3 & 1.060 & 5 &  0.0811 & 2.6 & 0.964 \\
 540.3 & 74.9 & 104 & 2.2 & 6 &  0.0746 & 2.4 & 1.063 & 5 &  0.0794 & 2.6 & 0.963 \\
 559.6 & 75.1 & 103 & 2.5 & 6 &  0.0741 & 2.6 & 1.077 & 5 &  0.0815 & 2.7 & 0.963 \\
\hline
\end{tabular}
\caption{\label{Franztab2} \small Fit results for $f^2$, obtained with the
standard Chew method ($f_{\rm c}^2 $), and with 
conformal mapping ($f_{\rm c, \,map}^2 $) for all energies with absolute normalization. 
$\Delta\theta $: angular range; $N$: number of data points; 
$s$: error of individual data point; $j_{\rm max}$: maximum polynomial 
order according to $F$ test;   
$ \Delta f^2 $: fit error 
of  $f_{\rm c}^2 $ and $f_{\rm c, \,map}^2 $, respectively; $k_{\rm corr}$: 
correction factor for systematic offset.}
\end{center}
\end{table*}
\subsection{\boldmath Determination of $f_c^2$ from our data}
Besides the systematic uncertainties caused by the errors of the 
data, there is also a dependence on the angular range used in the fit.
With decreasing range also the extracted coupling constant 
\begin{figure}[h]
\centerline{\epsfig{file=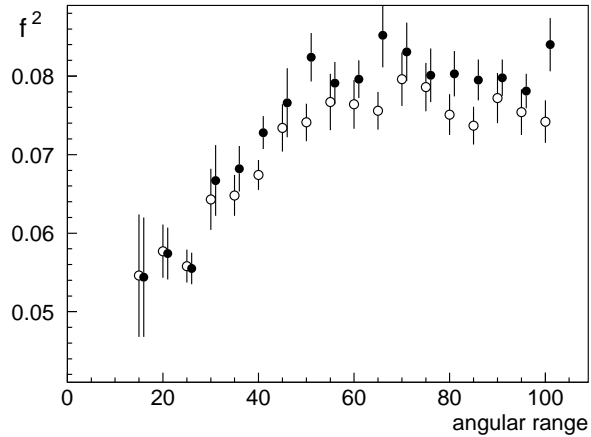,width=8.5cm}}
\caption{\label{Franzfig4} \small Dependence of $f^2$ fits on the angular range 
at $T_{\rm n} = 440$~MeV. Full dots are the results with conformal mapping.}
\end{figure}
is decreasing. This can be seen from the results on $f_{\rm c}^2$ 
and $f_{\rm c,\,map}^2$ 
obtained with the real data at $T_{\rm n} = 440$ MeV, as 
shown in Fig.~\ref{Franzfig4}.
For the fits at this energy all data points were given a statistical error 
of 2.8\,\%. With this choice the
reduced ${\chi}^2$ of the fit is of the order of 1 for the whole angular range. 
The error of  $f_{\rm c}^2$ is about 15\,\% 
for an angular range $\Delta \theta = 15^{\circ}$  and drops to about 3\,\% for 
$\Delta \theta = 100^{\circ}$.
According to Fig.~\ref{Franzfig4} an angular range $\Delta \theta 
\ge 50^{\circ}$ is needed. 
The variation of the extracted $f^2$ values is minimal for $75^{\circ} 
\le \Delta \theta \le 95^{\circ}$. 
For all energies it was found, 
that the angular range should extend to the region where the differential 
cross section reaches its 
minimum. In order to have no different weights from the four data sets, we 
have chosen for each laboratory kinetic energy a common relative error 
per data point, which was adjusted to get a 
reduced  $\chi_{\rm R}^2$ of the order of 1 for the whole angular range. 
The maximum polynomial order 
$j_{\rm max}$ was determined either when a minimum of $\chi_{\rm R}^2$ was reached  
or when  
$F 
\leq 4$,  
which corresponds to a significance of at least 95\,\% for the 
term with $j_{\rm max}$. The fit error of $f^2$
is then below 5\,\%.
The  results of the fits at energies 
where an absolute normalization exists 
are collected in Table~\ref{Franztab2}. 

The correction factors $k_{\rm corr}$, which were needed to correct for the 
systematic offset introduced by the Chew extrapolation were determined for 
both methods by an extended study with OPE pseudo data. 
For each energy we have generated 
1000 sets of pseudo data, with randomly distributed errors $s$ as given by the 
experimental data (see Table~\ref{Franztab2}). For the Chew method 
without conformal mapping no unique $j_{\rm max}$ was found for a given energy. 
Therefore we have used the $k_{\rm corr}$ obtained for that value of $j_{\rm max}$ 
which was obtained for the real data. This complication does not arise with 
conformal mapping, where the fit criteria (F-test) lead to 
the same value of $j_{\rm max}$ at all energies. Moreover, 
the energy variation of $k_{\rm corr}$ is only $\approx 1$\,\% as 
compared with almost 10\,\% without conformal mapping. 
Therefore we prefer to extract an energy independent value of  $f^2$ using the 
conformal mapping method. The maximum polynomial order $j_{\rm max}$, the 
resulting value of $f^2$, the fit error $\Delta f^2$  and the correction factor 
$k_{\rm corr}$ is given in Table~\ref{Franztab2} for both methods  
at each energy.

A contribution to the error which has not been taken into account so far is due to the 
uncertainty of $\approx 2\,\%$ in the relative normalization of data 
set IV with respect to the other sets. It was investigated systematically 
and a variation of about 1.5\,\% of the extracted $f^2$ was observed. 
This contribution is not contained in $\Delta f^2$ of Table~\ref{Franztab2}. 
It was, however, added in quadrature to each $\Delta f^2$  before the 
energy averaged weighted mean  
$\langle f^2_{\rm \pi np} \rangle = \langle f^2 \cdot k_{\rm corr}\rangle$ 
has been calculated.

For the prefered method with conformal mapping 
we  obtain as our final result the 
weighted mean of the charged coupling constant in the energy range $300\mbox{ MeV} 
\le T_{\rm n} \le 560\mbox{ MeV}$: 
\[\langle f_{\rm \pi np}^2 \rangle = 0.0760 
\pm 0.0008\;.\]
The given error is the propagated error calculated from the 
individual $\Delta f^2$. It is slightly larger than the 
standard deviation of the mean. 
 
It has to be kept in mind, that both, the relative 
error of the data points as well as the absolute normalization enter and 
limit the determination of $f^2$. In our case both error types are of the 
same order. With more precise data, i.e. smaller errors of the relative 
cross sections, the fitting procedure would be more stable 
and the systematic effects 
be smaller. Changes in the absolute normalization factor and its error, 
on the other hand, propagate to the $f^2$ values only with a 
factor $1/2$ since the Chew function at 
the pole depends on  $f^4$. 

All method tests and the calculation of $k_{\rm corr}$ have 
been performed with 
OPE model data. Therefore, a model contribution should eventually 
be added to the error margin. 

Our value of 0.0760 for the charged coupling constant is 
remarkably lower than the 
value obtained by H\"ohler et al.~\,\cite{Hoehler} from pion-nucleon scattering. 
It is, however, in good agreement with more recent determinations of 
$f^2$ from both, pion-nucleon partial wave analysis \,\cite{Arndt94} and 
nucleon-nucleon scattering \,\cite{fqArndt}.
\section*{Acknowledgments}
This work has been supported by the German Bundes\-mi\-ni\-ste\-rium  
f\"ur Bildung und Forschung.

\end{document}